\documentclass[12pt,a4paper,twoside,reqno]{amsart}
\usepackage{amsmath, amsthm, amssymb}
\usepackage{graphicx}

\DeclareMathOperator{\arctanh}{arctanh}
\title[Hyperboloidal approach to quasinormal modes]{A toy model of hyperboloidal approach to quasinormal modes}

\begin{document}

\author{Piotr Bizo\'n}
\address{Institute of Theoretical Physics, Jagiellonian University, Krak\'ow,
  Poland}
\email{bizon@th.if.uj.edu.pl}

\author{Tadeusz Chmaj}
\address{Institute of Nuclear Physics, Krak\'ow,
  Poland}
\email{tadeusz.chmaj@ifj.edu.pl}

\author{Patryk Mach}
\address{Institute of Theoretical Physics, Jagiellonian University, Krak\'ow,
  Poland}
\email{patryk.mach@uj.edu.pl}

\thanks{This research was supported by the Polish National Science
  Centre grant no.\ 2017/26/A/ST2/00530.}

\date{\today}

\begin{abstract}
We consider a scalar field propagating in the static region of the two dimensional de Sitter space. This simple system is used to illustrate the advantages of  hyperboloidal foliations in the analysis of quasinormal modes.
\end{abstract}
\maketitle

\section{Introduction}
Many physical systems respond to perturbations by oscillating at certain characteristic frequencies. For closed systems (such as a guitar string) these frequencies, called normal frequencies, are real and correspond to the eigenvalues of a self-adjoint operator. For open systems (such as waves scattering off an obstacle or a black hole), the characteristic frequencies, called quasinormal (or scattering) frequencies, are complex. An imaginary part of the quasinormal frequency determines the exponential decay of the amplitude of oscillation, which is due to the loss of energy by radiation. In the literature, quasinormal modes are traditionally defined by imposing  an outgoing wave condition at infinity \cite{lp} (and, in the case of black holes, also an ingoing boundary condition at the horizon \cite{chandra}), which implements the physical condition that nothing is `coming in from infinity' (or from the horizon). While in most  situations this definition works fine, it is not quite satisfactory both from the mathematical and physical viewpoints (for an excellent  discussion of this issue, see \cite{w1}). The problem is that the unitary evolution,  based on the standard constant time foliations of spacetime,  does not provide a natural setting for understanding the  dissipation-by-dispersion phenomena. This drawback can be remedied  by using hyperboloidal foliations and the associated non-unitary evolution which inherently incorporates the loss  of energy by radiation. In this formulation the quasinormal modes can be defined as genuine eigenmodes of a certain non self-adjoint linear operator.  To our knowledge, the  hyperboloidal approach to  quasinormal modes was first suggested  by  Schmidt \cite{bernd}. In the past decade this idea has been implemented numerically \cite{brz, mja} and developed rigorously  in the mathematical literature \cite{w1, sd, gw}; it also featured in the recent proof of non-linear stability of the Kerr-de Sitter black holes by Hintz and Vasy \cite{hv}.

The purpose of this pedagogical note, addressed to physicists, is to present a very simple toy model illustrating the advantages of the  hyperboloidal approach to quasinormal modes.

\section{Setup}
 Consider a two-dimensional manifold $\mathcal{M} = \{t \in \mathbb R, x \in (-1,1)\}$ with the metric
\begin{equation}
\label{g5fd}
g = -(1 - x^2) dt^2 + (1 - x^2)^{-1} dx^2.
\end{equation}
It corresponds to the static region of the two dimensional de Sitter spacetime with constant scalar curvature $R(g) = 2$. We introduce the hyperboloidal foliation
\[ \tau = t + \frac{1}{2} \log(1 - x^2). \]
In terms of the coordinates $(\tau,x)$ the metric (\ref{g5fd}) takes the form
\begin{equation}\label{metric2}
 g = -(1 - x^2) d\tau^2 - 2 x d\tau dx + dx^2,
 \end{equation}
which is regular at the cosmological horizons $x = \pm 1$.

We are interested in the propagation of a  scalar field with mass $m$ on $\mathcal{M}$, as described by the Klein-Gordon equation
\begin{equation}\label{kg}
(\Box_g-m^2) \phi = 0,
\end{equation}
where $\Box_g \equiv g^{\alpha\beta} \nabla_\alpha \nabla_\beta$ and $\nabla_\alpha$ denotes the covariant derivative with respect to the metric $g$.
In terms of the coordinates $(\tau,x)$ we have
\begin{equation}
\label{we}
-\partial_{\tau \tau} \phi - 2 x \partial_{\tau x} \phi - \partial_\tau \phi + \partial_x \left[ (1 - x^2) \partial_x \phi \right] - m^2 \phi = 0,
\end{equation}
\begin{equation}
\label{ic}
\phi(0,x) = f(x), \quad \phi_\tau(0,x) = g(x).
\end{equation}
We assume that the functions $f(x)$ and $g(x)$ are smooth  on $x \in [-1,1]$. Note that the curves $x = \pm 1$ are null; consequently no boundary conditions are imposed at the endpoints $x = \pm 1$.

Multiplying equation \eqref{we} by $\partial_\tau \phi$, one gets the conservation law
\begin{equation}
\label{cca1}
\partial_\tau \rho + \partial_x j = 0,
\end{equation}
where
\[ \rho = \frac{1}{2}\left[ (\partial_\tau \phi)^2 + (1 - x^2) (\partial_x \phi)^2 +m^2 \phi^2 \right], \]
and
\[ j = x (\partial_\tau \phi)^2 - (1 - x^2) \partial_\tau \phi \partial_x \phi. \]
Integrating the conservation law (\ref{cca1}) over the curve $\tau = \mathrm{const}$ and defining the Bondi energy $\mathcal E = \int_{-1}^1 \rho dx$, one gets
\[ \frac{d\mathcal E}{d\tau} = - \left[ \partial_\tau \phi (\tau,1) \right]^2 -\left[ \partial_\tau \phi (\tau,-1) \right]^2, \]
which shows that the Bondi energy $\mathcal E$ decreases due the fluxes of outgoing radiation across the cosmological horizons. This indicates that for $\tau\rightarrow \infty$ the solution  tends to a static equilibrium, which in case of \eqref{we} is just a constant (equal to zero if $m$ is nonzero).

In the following we first discuss the case $m=0$, which can be solved explicitly, and then analyze the case $m\neq 0$ using the Galerkin method.

\section{Massless scalar field}

In this section we set $m=0$ in equation \eqref{we}. Then, in terms of double null coordinates (see Fig.~1)
\begin{equation}
\label{uv}
U = (1 - x)e^{-\tau}, \quad V = (1 + x)e^{-\tau},
\end{equation}
equation (\ref{we}) is equivalent to $ \partial_{U V} \phi  = 0$. Hence the general solution can be written as
\begin{equation}\label{general} \phi(\tau,x) = \phi_1(U) + \phi_2(V), \end{equation}
where $\phi_1$ and $\phi_2$ are arbitrary  functions. As a consequence, the solution of equation \eqref{we} with initial data \eqref{ic} can be represented by  the d'Alembert formula, which is the case at hand takes the form
\[ \phi(\tau,x)  =  (1 - \frac{1}{2} U) f( 1 - U) + (1 - \frac{1}{2} V) f(V - 1) + \frac{1}{2} \int_{V - 1}^{1 - U} \left[ g(z) - f(z) \right] dz. \]
It follows immediately from the above formula that the end-state of the evolution is a constant given by
\begin{equation}
\label{endstate}
\phi_\infty \equiv \lim_{\tau\rightarrow \infty} \phi(\tau,x) =  f(1) + f(-1) + \frac{1}{2} \int_{-1}^1 \left[g(z) - f(z)  \right] dz .
\end{equation}

\begin{figure}[t]
\includegraphics[width=.5\textwidth]{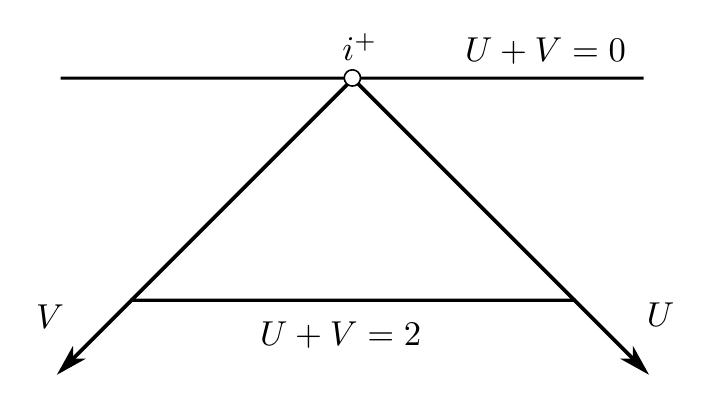}
\caption{Conformal diagram for metric (\ref{g5fd}).}
\end{figure}

\noindent In order to see the rate of convergence to the end-state, let us expand \eqref{general} in the Taylor series around $e^{-\tau} = 0$. We get
\begin{eqnarray}
\phi (\tau, x)  & = & \phi_1 \left[ (1-x) e^{-\tau} \right] + \phi_2 \left[ (1+x) e^{-\tau} \right] \nonumber\\
& = & \phi_1(0) + \phi_2(0) + \phi_1^\prime(0) (1 - x) e^{-\tau} + \phi_2^\prime(0) (1 + x) e^{-\tau} \nonumber\\
& & + \frac{1}{2} \phi_1^{\prime \prime}(0) (1 - x)^2 e^{-2\tau} + \frac{1}{2} \phi_2^{\prime \prime}(0) (1 + x)^2 e^{-2 \tau} + \dots \label{expansion}
\end{eqnarray}
The terms decaying as $e^{-n\tau}$ in this  expansion correspond to quasinormal modes. To see this, let us insert the ansatz $\phi(\tau,x) = e^{\lambda \tau} \psi(x)$ into equation \eqref{we}. This yields the quadratic eigenvalue problem
\begin{equation} \label{eigen} (1-x^2) \psi^{\prime \prime} - 2x(\lambda + 1) \psi^\prime - \lambda(\lambda + 1)  \psi = 0,
\end{equation}
whose general solution is
\[ \psi(x) = A \, (1+x)^{-\lambda} +  B \, (1-x)^{-\lambda}, \]
where $A$ and $B$ are constants. The quasinormal modes are defined as smooth solutions of equation \eqref{eigen}. The requirement of smoothness implies quantization of eigenvalues $\lambda_n = -n$, $n=0, 1, 2, \dots$ For $\lambda_0 = 0$ the eigenfunction is constant $\psi_0=1$, while for $n\geq 1$ there is a two-fold degeneracy
\[ \psi_n^{\pm}(x) = (1 + x)^{n} \pm   (1-x)^{n}, \]
where for convenience we took odd and even combinations.

Returning to the expansion \eqref{expansion} and defining $\phi_1(0) = \frac{1}{2}(a_0 - b_0)$, $\phi_2(0) = \frac{1}{2}(a_0 + b_0)$, $\phi^\prime_1(0) = a_1 - b_1$, $\phi^\prime_2(0) = a_1 + b_1$, $\phi^{\prime \prime}_1(0) = 2(a_2 - b_2)$, $\phi^{\prime \prime}_2(0) = 2(a_2 + b_2)$, \dots, we obtain the late-time behavior as a superposition of quasinormal modes
\begin{eqnarray*}
\phi (\tau, x)  & =& a_0 + \left[ a_1 \psi_1^+(x) + b_1 \psi_1^-(x) \right] e^{-\tau} + \\
& & + \left[ a_2 \psi_2^+(x) + b_2 \psi_2^-(x) \right] e^{-2\tau} + \dots
\end{eqnarray*}
It is clear from this expression that no quasinormal modes are excited for the initial data $f$ and $g$ that are  supported away from the cosmological horizons $x=\pm 1$.

\vskip 1em

\noindent \emph{Remark.} It is instructive to compare the above approach with the traditional definition of quasinormal modes as outgoing wave solutions.
 Using time $t$ and defining the tortoise coordinate $y = \arctanh{x}$, one transforms equation \eqref{kg} with $m=0$ into $\partial_{tt} \phi=\partial_{yy} \phi$. Separating time $\phi(t,y)=e^{\lambda t} \xi(y)$, we get
$ \xi''-\lambda^2 \xi=0,$
whose general solution is
$ \xi(y)= A e^{\lambda y} + B e^{-\lambda y}\,$.
Thus, there are no solutions that are outgoing both at $y=-\infty$ and $y=\infty$.
Note that in the coordinates $(t,y)$
the quasinormal modes $e^{-n \tau} \psi_n^{\pm}(x)$ are given by
$ e^{-n (t - y)} \pm e^{-n (t + y)}$.

\section{Massive scalar field}
Since for $m \neq 0$ there is no d'Alembert formula available, we shall use a different approach, based on the Galerkin method,  similar to the one  developed by two of us  in \cite{bm}. This will allow us to solve the initial value problem explicitly for all polynomial initial data.

We begin by expanding the solution in Legendre polynomials $P_n(x)$
\begin{equation}
\label{aaa3}
\phi(\tau,x) = \sum_{n=0}^\infty a_n(\tau) P_n(x).
\end{equation}
Inserting this series into equation \eqref{we} we obtain the infinite system of ordinary differential equations for the coefficients $a_n(\tau)$
\begin{equation}\label{an}
\ddot a_n + (2n+1) \dot a_n + [n(n+1) + m^2] a_n + 2(2n+1) f_n=0,
\end{equation}
where the dot denotes differentiation with respect to $\tau$  and we defined
\begin{equation}\label{fn}
f_n(\tau) \equiv \dot a_{n+2}(\tau) + \dot a_{n+4}(\tau) + \dots.
\end{equation}
In  deriving \eqref{an}, the projection of the term $2x \partial_{\tau x} \phi$ was obtained using the identity
\begin{eqnarray*}
 x  P_n'(x) & = &  n P_n(x) + \sum_{n-2j\geq 0} (2n+1-4j) P_{n-2j}(x),
\end{eqnarray*}
which follows readily from  Bonnet's recursion formula
\[ (2n+1) x P_n(x) = (n+1) P_{n+1}(x) + n P_{n-1}(x).  \]

\noindent Note that the even and odd modes decouple in the system \eqref{an}.
\vskip 0.1cm
  Assuming that $f_n(\tau)$ is known, one can formally solve equation \eqref{an}
as an inhomogeneous linear equation with constant coefficients. The characteristic equation for the homogeneous part is
\[r^2 +(2n+1) r +[n(n+1) + m^2]=0,\]
which has roots
\begin{equation}\label{roots}
  r=- n -\frac{1}{2}\pm \frac{1}{2} \sqrt{1 - 4 m^2},
\end{equation}
hence there are three different cases depending  on whether $m^2$ is less, equal, or larger than $1/4$. To illustrate the method let us consider the case $m^2>1/4$ (the other two cases can be treated in an analogous way).
In this case the solution of the homogeneous part of equation \eqref{an} has the form
\begin{equation}\label{hom}
a_n^{\mathrm{hom}}(\tau) = e^{-\gamma_n\tau} \left[A_n \cos\left(\omega \tau \right) + B_n \sin\left(\omega\tau \right)\right],
\end{equation}
where $\gamma_n=n+\frac{1}{2}$, $\omega=\frac{1}{2} \sqrt{4 m^2 - 1}$ and $A_n, B_n$ are constants. Using the method of variations of constants one can write the solution of equation \eqref{an} for a given $f_n$ as
\begin{align}\label{inhom}
& a_n(\tau)  =  a_n^{\mathrm{hom}}(\tau) \nonumber  \\
&  + \frac{4\gamma_n}{\omega} e^{-\gamma_n\tau} \cos\left( \omega\tau \right) \int_0^\tau \sin\left(\omega\tau^\prime \right) e^{\gamma_n\tau^\prime} f_n(\tau^\prime) d \tau^\prime \nonumber\\
&-\frac{4\gamma_n}{\omega} e^{-\gamma_n\tau}  \sin\left( \omega\tau \right) \int_0^\tau \cos\left(\omega \tau^\prime \right) e^{\gamma_n \tau^\prime} f_n(\tau^\prime) d \tau^\prime.
\end{align}
The constants $A_n$ and $B_n$ can be expressed in terms of the initial values of $a_n$ and $\dot a_n$ as follows
\begin{equation}\label{icond}
A_n = a_n(0), \quad \omega B_n = \dot a_n(0) + \gamma_n a_n(0).
\end{equation}

Suppose now that the initial data are polynomial in $x$, hence they consist of a finite number of Legendre modes. Let the highest mode number be $N$. The solutions for $a_N$ and  $a_{N-1}$ are then given by the solutions of the homogeneous equation \eqref{hom}. Once these solutions are known, one can easily get solutions for  $a_{N-2}$ and $a_{N-3}$ by a straightforward application of formula \eqref{inhom}. This procedure can then be iterated backwards, until the solutions for $a_0$ and $a_1$ are obtained. For example, for $N = 2$ we have  from \eqref{hom} and \eqref{icond}
\begin{eqnarray*}
a_2(\tau) & = &  e^{-5 \tau /2} \left[\frac{5 a_2(0)+2 \dot{a}_2(0)}{2\omega} \sin \left(\omega\tau \right)+ a_2(0) \cos \left(\omega\tau \right)\right], \\
a_1(\tau) & = & e^{-3 \tau /2} \left[\frac{3 a_1(0)+2 \dot{a}_1(0)}{2\omega} \sin \left(\omega\tau \right)
 + a_1(0) \cos \left(\omega \tau \right)\right],
\end{eqnarray*}
and, next, inserting $f_0(\tau)=\dot a_2(\tau)$ into formula \eqref{inhom} we get
\[
a_0(\tau) = e^{-\tau /2} \left[c_1 \sin(\omega\tau)+c_2  \cos(\omega\tau)  \right]
 +e^{-5\tau /2} \left[c_3 \sin(\omega\tau)+c_4 \cos(\omega\tau)  \right],
\]
where
\begin{eqnarray*}
c_1 & =& \frac{4 (\omega^2 + 1)[a_0(0) + 2 \dot a_0(0)] + (4 \omega^2 + 25) a_2(0) - (4 \omega^2 - 2) \dot a_2(0)}{8 \omega (\omega^2 + 1)}, \\
c_2 & =& \frac{8 (\omega^2 + 1) a_0(0) - (4 \omega^2 + 25) a_2(0) - 6 \dot a_2(0)}{8 (\omega^2 + 1)}, \\
c_3 & =& \frac{ (4\omega^2 + 25) a_2(0) + 2 (2\omega^2 + 5) \dot a_2(0)}{8 \omega (\omega ^2 + 1)}, \\
c_4 & =& \frac{(4\omega^2 + 25) a_2(0) + 6 \dot a_2(0)}{8(\omega ^2 + 1)}.
\end{eqnarray*}
\vskip 0.2cm
The quasinormal modes can be determined as in the massless case by the separation of variables  $\phi(\tau,x) = e^{\lambda \tau} \psi(x)$ which, when inserted into \eqref{we}, leads to the quadratic eigenvalue problem
\begin{equation} \label{eigen2} (1-x^2) \psi^{\prime \prime} - 2x(\lambda + 1) \psi^\prime - [\lambda(\lambda + 1)+m^2]  \psi = 0.
\end{equation}
Searching for  solutions  in the form of a power series
$\psi(x) = \sum_{n=0}^\infty b_n x^n$ gives
the recurrence relation
\[ b_{n+2} = \frac{\lambda^2 + (1 + 2n)\lambda + n(n+1) + m^2}{(n+1)(n+2)}\, b_n\,.\]
For the solution to be smooth, it is necessary that the power series terminates at some finite $n$; this happens for
\begin{equation}\label{lambdan}
\lambda =\lambda_n\equiv  - \frac{1}{2} \left( 2n + 1 \pm \sqrt{1 - 4 m^2} \right), \quad n = 0, 1, \dots
\end{equation}
The corresponding eigenfunctions $\psi_n(x)$ are then polynomials (but not orthogonal ones).
As follows from \eqref{lambdan}, for $m^2\leq 1/2$ the quasinormal modes are purely damped, while for $m^2>1/4$ they are damped oscillations.


\begin{thebibliography}{99}
\bibitem{lp} P.\ D.\ Lax, R.\ S.\ Phillips, \emph{Scattering Theory}, Academic Press, New York 1967
\bibitem{chandra} S.\ Chandrasekhar, \emph{The Mathematical Theory of Black Holes,} Clarendon Press, Oxford 1983
\bibitem{w1} C.\ M.\ Warnick, \emph{On quasinormal modes of asymptotically anti-de Sitter black holes,} Commun.\ Math.\ Phys.\ 333, 959 (2015)
\bibitem{bernd} B.\ G.\ Schmidt, \emph{On relativistic stellar oscillations,} Gravity Research Foundation essay (1993)
\bibitem{brz} P.\ Bizo\'n, A.\ Rostworowski, A.\ Zengino\u{g}lu, \textit{Saddle-point dynamics of a Yang-Mills field on the exterior Schwarzschild spacetime}, Class.\ Quantum Grav.\ 27, 175003 (2010)
\bibitem{mja}  R.\ P.\ Macedo, J.\ L.\ Jaramillo, M.\ Ansorg, \emph{Hyperboloidal slicing approach to quasi-normal mode expansions: the Reissner-Nordstr\"om case,} Phys. Rev. D 98, 124005 (2018)
\bibitem{sd} S.\ Dyatlov, \emph{Quasi-normal modes and exponential energy decay for the Kerr-de Sitter black hole,} Commun.\ Math.\ Phys.\ 306, 119 (2011)
\bibitem{gw} D.\ Gajic, C.\ M.\ Warnick, \emph{Quasinormal modes in extremal Reissner-Nordstr\"om spacetimes,} arXiv:1910.08479
\bibitem{hv} P.\ Hintz, A.\ Vasy, \emph{The global non-linear stability of the Kerr-de Sitter family of black holes,} Acta  Mathematica 220, 1 (2018)
\bibitem{bm} P.\ Bizo\'{n}, P.\ Mach, \emph{Global dynamics of a Yang-Mills field on an asymptotically hyperbolic space}, Trans.\ Amer.\ Math.\ Soc. 369, 2029 (2017)
\end{thebibliography}
\end{document}